\documentclass{article}
 \usepackage[pdftex]{graphicx} 
\usepackage{latexsym}
\usepackage{epstopdf}
\usepackage{float}
\usepackage{amssymb}
\usepackage{amsmath}
\usepackage{color}
\usepackage{url}
\usepackage{booktabs}
\usepackage[bottom]{footmisc}

\usepackage{amsmath}
\usepackage{amssymb}
\usepackage{color}
\usepackage{graphicx}
\usepackage{fullpage}
\usepackage{subfigure}
\usepackage{lscape}
\usepackage{epstopdf}
\usepackage{hyperref}

\newcommand{\etal}{{\it{}et~al.}}

\usepackage{setspace}

\title{Urban characteristics attributable to density-driven tie formation\footnote{This work is appeared in NetSci 2012, Evaston, IL, June 20th, 2012.}}
\author{Wei Pan,$^{1}$ Gourab Ghoshal,$^{2}$ Coco Krumme,$^{1}$ Manuel Cebrian,$^{1,3,4}$ and Alex Pentland$^{1}$\footnote{Corresponding author: pentland@mit.edu.} \\
\\
\normalsize{$^{1}$Media Laboratory, Massachusetts Institute of Technology, Cambridge, MA, USA}\\
\normalsize{$^{2}$Department of Earth and Planetary Sciences, Harvard University, Cambridge, MA, USA}\\
\normalsize{$^{3}$Department of Computer Science and Engineering, UC San Diego, La Jolla, CA, USA}\\
\normalsize{$^{4}$National Information and Communications Technology Australia, Melbourne, VIC, Australia}\\
}
\date{}

\begin{document}

\maketitle

\begin{abstract}
{\bf Motivated by empirical evidence on the interplay between geography, population density and societal interaction, we propose a generative process for the evolution of social structure in cities. Our analytical and simulation results predict both super-linear scaling of social tie density and information flow as a function of the population.  We demonstrate that our model provides a robust and accurate fit for the dependency of city characteristics with city size, ranging from  individual-level dyadic interactions (number of acquaintances, volume of communication) to population-level variables (contagious disease rates, patenting activity, economic productivity and crime) without the need to appeal to modularity, specialization, or hierarchy.}

\end{abstract}

\vspace*{0.5cm}

A larger percentage of people live in cities than at any point in human history~\cite{crane2005nature}, while the density of urban areas is generally increasing~\cite{censuspop}. One of the enduring paradoxes of urban economics concerns why people continue to move to cities, despite elevated levels of crime, pollution, and wage premiums that have steadily lost ground to premiums on rent~\cite{glaeser2000consumer}. New York in the 18th century, according to Thomas Jefferson, was ``a toilet of all the depravities of human nature''. Since Jefferson's day, the city has grown to host the depravities of 100-fold more people, yet the stream of new arrivals has not stemmed.  

While the forces behind any urban migration are complex, the advantages afforded by urban density comprise an important driver. Smith~\cite{smithwealth} was one of the first to point to urban centers as exceptional aggregators, whether of innovations or depravities. Cities appear to support levels of enterprise impossible in the countryside, and urban areas use resources more efficiently, producing more patents and inventions with fewer roads and services per capita than rural areas~\cite{milgram1970experience, becker1999population, krugman1993number, masahisa1999spatial, bettencourt2007growth,bettencourt2010unified}.  


Despite the scientific interest these patterns have generated, we still lack a compelling generative model for why an agglomeration of people might lead to the more efficient creation of ideas and increased productivity in urban areas. Models of agglomeration point to the role of technology diffusion in creating intellectual capital \cite{audretsch1996r,jaffe1993geographic, anselin1997local}, but lack a quantitative description of the generative mechanism.  Hierarchies have also been proposed as an elegant mechanism for this growth~\cite{arbesman2009superlinear}; however, recent studies hint at the absence of well-defined hierarchy across geographical scales \cite{leskovec2009community,expert2011uncovering,onnela2011geographic,ahn2010link,mucha2010community}.

Recent developments in the study of social networks shed some light on this challenge. Empirical evidence suggests that interactions and information exchange on social networks are often the driving force for idea creation, productivity and  and individual prosperity. Examples of this include the theory of  weak ties~\cite{granovetter1973strength,granovetter2005impact}, structural holes~\cite{burt1995structural}, the strong effect of social interaction on economic and social success ~\cite{eagle2010network}, the influence of face-to-face interactions on the effect of productivity~\cite{wu2008mining}, as well as the importance of information flow in the management of  Research and Development~\cite{allen2003managing,reagans2001networks}. Consequently,  it seems that understanding the mechanism of tie formation in cities is the key to the development of a general theory for a city's growth described by it's economic indicators and its population.

In this paper we present a simple, bottom-up, robust model describing the efficient creation of ideas and increased productivity in cities. Our model can be regarded as a natural extension of Krugman's insights on industries~\cite{krugman1993number}. Krugman pointed out the connection between manufacturing efficiency and transportation of goods as a function of proximity of factories. Similarly, our theory connects the efficiency of idea-creation and information flow to the proximity of individuals generating them.  

Our model consists of two essential features.  We propose a simple analytical model for the number of social ties $T(\rho)$ formed between individuals, with population density $\rho$ as its single parameter.  We demonstrate that increases in density and proximity of populations in cities leads to a higher \emph{social tie density} for urban population. We then show that the diffusion rate along these ties---a proxy for the amount of information flow---accurately reproduces the empirically measured scaling of urban features such as rate of AIDS infections, communication and GDP.  

The model naturally leads to a super-linear scaling of indicators with city population~\cite{bettencourt2007growth} without the need to resort to any parameter-tuning (although it predicts a different functional form than a simple power-law and is a more accurate match to the data). The surprisingly similar scaling exponent across many different urban indicators (see Supplementary Information, Section S1), suggests a common mechanism  behind them.  Social tie density and information flow therefore offer a parsimonious, generative link between human communication patterns, human mobility patterns, and the characteristics of urban economies, without the need to appeal to hierarchy, specialization, or similar social constructs.



\vspace*{10px}
\noindent \textbf{A model for social-tie density}
\vspace*{10px}

\noindent We propose to model the formation of ties between individuals (represented as nodes) at the resolution of urban centers. Since our model is based on geography, a natural setting for it is a 2D Euclidean space with nodes denoted by the coordinates $\vec{x}_i \in {\mathbf R}^2$ on the infinite plane.  Furthermore, we also assume that these nodes are distributed uniformly in space, according to a density $\rho$ defined as,
\[\rho = \text{\# nodes per unit area}. \] 
While the assumption of uniform density is an approximation, the qualitative features of the model are unaffected by other more realistic choices of the density distribution---see Supplementary Information, Section S5.  Following~\cite{liben2005geographic}, 
we define the probability of a tie to form between two nodes $i,j$ in the plane as
\begin{equation}
P_{ij} \propto \frac{1}{\mathrm{rank}_i (j)},
\label{eq:probij}
\end{equation}
where the rank is defined as
\begin{equation}
\mathrm{rank_i (j)}:= \left|\{k:d(i,k) < d(i,j)\}\right|,
\label{eq:rank}
\end{equation}
and $d_{ij}$ is the Euclidean distance between two nodes.
If $j$ lies at a radial distance $r$ from node $i$, then the number of neighbors closer to $i$ than $j$ is the product of the density and the area of the circle of radius $r$, and thus the rank is simply,
\begin{equation}
\mathrm{rank_i}(j) = \rho \pi r^{2},
\label{eq:rankcont}
\end{equation}
which implies that the probability an individual forms a tie at distance $r$ goes as $P(r) \sim 1/\pi r^2$, similar in spirit to a gravity model~\cite{krings2009gravity}.

For a randomly chosen node, integrating over $r$ up to an urban mobility ``boundary" denoted as $r_{max}$,  we obtain the expected number of social ties $ t(\rho)$. \begin{eqnarray}
t(\rho)  &=& \ln{\rho} + C \label{i2}, 
\end{eqnarray}
where $C = 2\ln{r_{max}} + \ln{\pi} + 1$.  We note that $r_{max}$ may well be unique for each city, and is often determined by geographical constraints as well as city infrastructure (cf. Supplementary Information, Section S3).
Integrating over the number of social ties for all nodes within an unit area gives us the social tie density $ T(\rho)$,
\begin{eqnarray}
T(\rho) &=& \rho\ln{\rho} + C' \rho,
\label{eq:tprime}
\end{eqnarray} 
with $C'=C-1$.  Thus the density of social ties formed between individuals grows as $T(\rho) \sim \rho \ln \rho$, a super-linear scaling consistent with the observations made by Calabrese \etal~\cite{calabrese2011connected}.
We argue that $T(\rho)$ to a first approximation is the  individual dyadic-level ingredient behind the empirically observed growth of city indicators. For more detail on the theoretical analysis and support for the assumptions involved, see Supplementary Information, Sections S3-S5. 

In order to test this theoretical result, we perform simulations of tie formation with more realistic discrete settings. Urban areas differ dramatically in both regional boundaries and population density. It is thus important to test the sensitivity of the model to a diversity of input parameters for the density $\rho$ and the urban ``boundary" $r_{max}$. We start from an empty lattice of size $N \times N$, with $N^2$ possible locations. 
The density $\rho$ is gradually increased by randomly assigning
new nodes to empty locations on the grid, where each node represents a small community, or city block of $10^2$ individuals. Once a node is added, 
the probability of forming a tie with one of its existing neighbors is computed by counting
the number of nodes closer to this node according to Eq.~\ref{eq:probij}. 
To test the sensitivity of our results to the relevant parameters we vary the size of the grid 
$(20 \le N \le 400)$ to mimic different scales for city boundaries $r_{max}$.  In addition we also vary city population between $10^4$ to $10^7$ residents as well as the functional form of the density distribution.  

\begin{landscape}
\begin{figure}
\centering
\includegraphics[width=23cm]{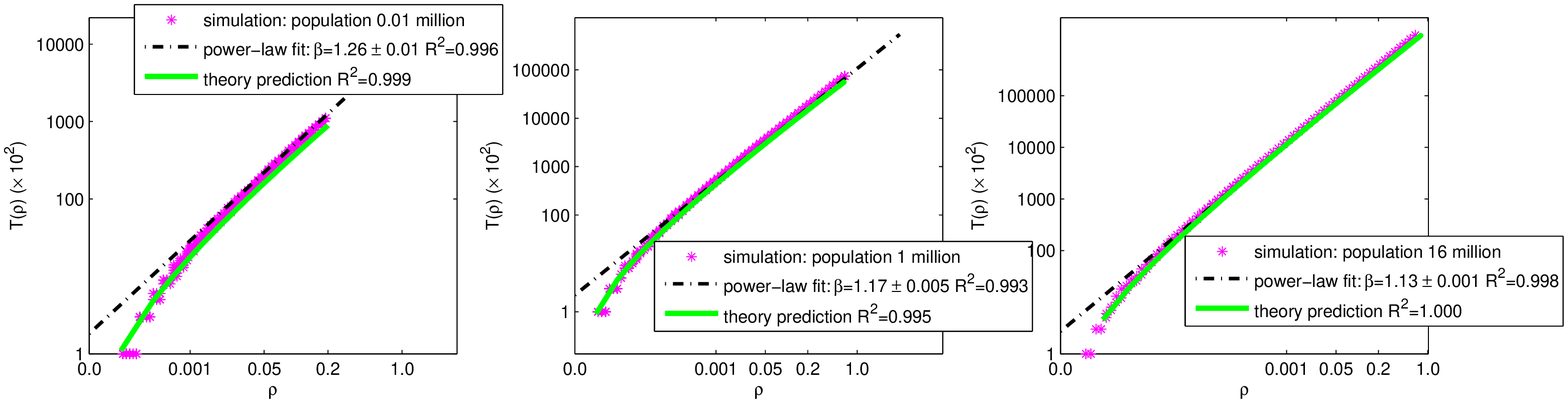}
\caption{The number of ties $T(\rho)$ plotted as a function of $\rho$ for various grid sizes $N$ and ``mobility limits" $r_{max}$. The solid green line represent the average over 30 realizations of the simulation described in the text, the solid line is the theoretical expression~\eqref{eq:tprime} while the dotted line is a fit to the form $T(\rho) \sim \rho^{\beta}$. As can be seen in each case the agreement between theory and simulation is excellent. The best fit to the scaling exponent yields a value of $\beta \approx 1.15$ independent of $N$. Note that the measured value of the exponent in empirical data is $1.1 \le \beta \le 1.3$.}
\label{fig:interaction}
\end{figure}
\end{landscape}

In Fig.~\ref{fig:interaction} we show the average over $30$ realizations of the simulation for different values of the grid size $N$ and city boundary $r_{max}$. The density $\rho$ in this case represents
the relative percentage of occupied locations on the grid, and $T(\rho)$  the total number of ties formed between nodes.  As Fig.~\ref{fig:interaction}  shows, the agreement between the theoretical expression for $T(\rho)$ \eqref{eq:tprime} and the curves generated by the simulation, is excellent at all scales despite our continuum approximation ($R^2 \approx 1$). 


As a comparative exercise, on the same plot, we also show the best fit to the form $T(\rho) \sim \rho^{\beta}$ and find a value of $\beta \approx 1.16$.   We note that this value is strikingly similar to empirically observed values by fitting a power law to the relationship between population and urban indicators.  It has been suggested that a fit of the form
$x \ln x$ can easily be mistaken for $x^\beta$ \cite{shalizi}, which together with our model suggests that observed scaling of cities may alternatively be described by Eq.~\eqref{eq:tprime}.  The latter functional form is additionally supported by the fact that it represents a generative model for the emergence of urban features as a result of density-driven communication patterns, without any parameter tuning or a priori assumption about the structure of the underlying social network.  Our simulation results indicate that  the scaling described in Eq.~\eqref{eq:tprime} is robust with respect to the choice of  different functional forms for the density distribution. (Supplementary Information, Section S5).

\vspace*{10px}
\noindent \textbf{Results}
\vspace*{10px}

\begin{figure}[h]
\centering
\includegraphics[width=\textwidth]{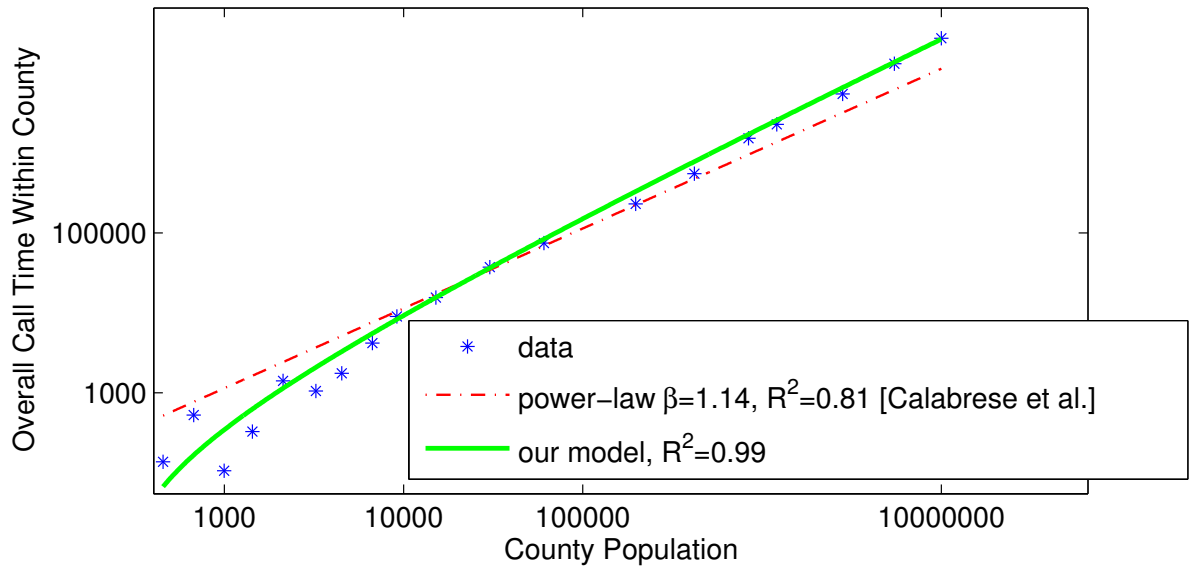}
\caption{Aggregated call time data {\it vs.} different densities (adapted from Calabrese \etal~\cite{calabrese2011connected}) together with the theoretical prediction from our model which connects density $\rho$ and number of social ties $T(\rho)$ via the $\rho \log \rho$ function. Our model captures both the power-law growth pattern and tilts on both end of the growth curve ($R^2=0.99$ vs. $R^2=0.81$, and provides a better fit for the data than the power-law model. }
\label{fig:ours}
\end{figure}

\paragraph{Empirical evidence for the effect of social tie density} Recent work~\cite{calabrese2011connected} shows a super-linear relationship between calling volume (time) and population across different counties in the United States.  As Fig. \ref{fig:ours} illustrates, the super-linear relationship in the data is approximated by the authors as a power-law growth $y= ax^{\beta}$ with $\beta\approx1.14$.  However, by assuming a uniform distribution on county sizes and treating population as a proxy for density, we show that our density driven model is able to capture precisely the distribution of the call volume. The model produces the exact shape of the curve, including the power-law growth pattern ($\beta = 1.14$) and tilts on both end, with an adjusted $R^2=0.99$ (See Fig. \ref{fig:ours}). Consequently, we propose that the model may well provide a reasonable explanation for communication patterns observed in US counties. 

\paragraph{Information diffusion with social-tie density} We note that the expected pattern of link formation in itself is insufficient to explain how growth processes in cities work to create observed scaling phenomena.  Instead the manner in which these links spread information determines value-creation and productivity.  Since it is known that social network structure has a dramatic effect on the access of information and ideas \cite{granovetter1973strength,eagle2010network, burt1995structural,wu2008mining,allen2003managing,reagans2001networks}, it seems plausible that higher social tie density should engender greater levels of information flow and interaction leading to the observed increases in productivity and innovation. 



\begin{landscape}
\begin{figure}[h]
\centering
\subfigure{
\includegraphics[width=0.5\textwidth]{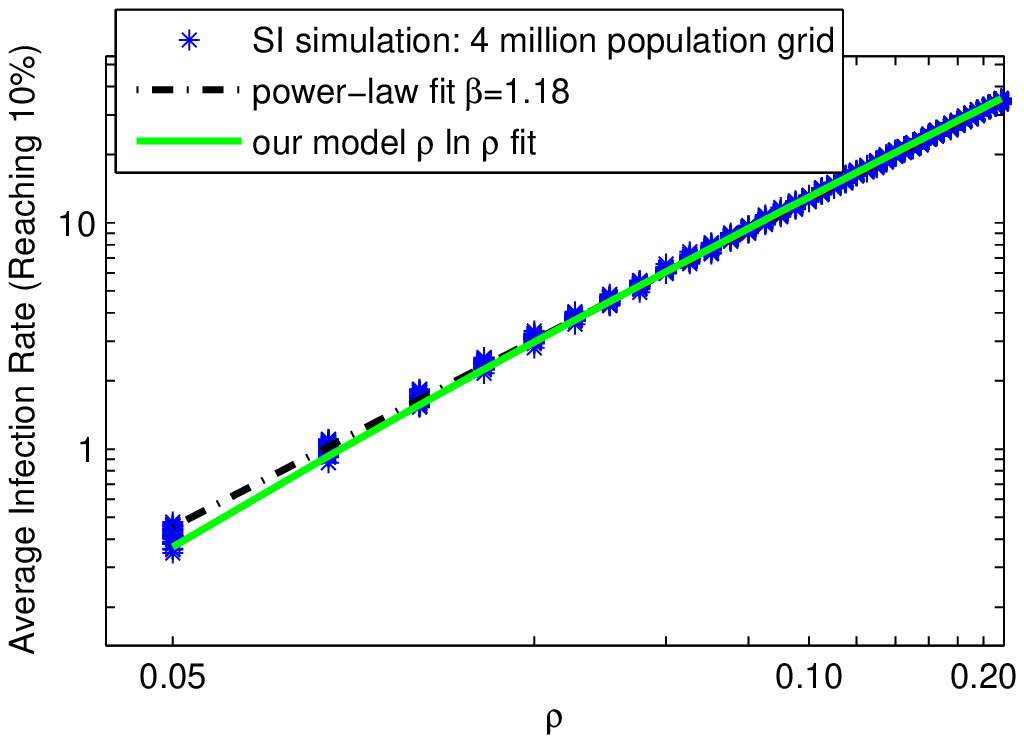}
}
\subfigure{
\includegraphics[width=0.5\textwidth]{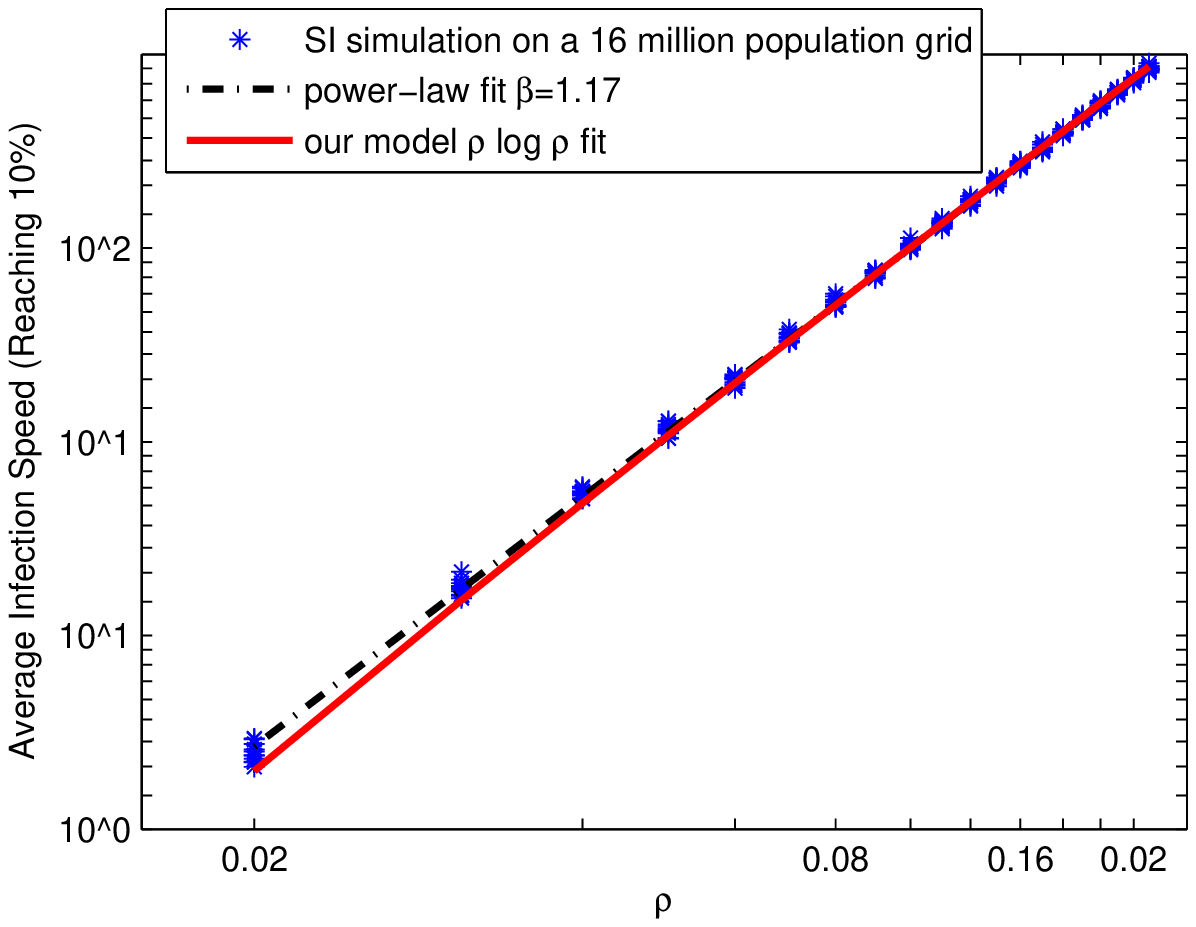}
}

\caption{{\footnotesize {\bf a)} The mean spreading rate as a function of density $\rho$ . The points correspond to a average over 30 realizations of simulations of the SI model on a $200\times200$ grid. The dashed line corresponds to a fit of the form $R(\rho) \sim \rho^{1+\alpha}$ with $\alpha = 0.18$. The solid line is a fit to our social-tie density model. {\bf b)} The mean spreading rate as a function of $\rho$  under the complex contagion diffusion model. The dashed line corresponds to the power-law fit of the form $R(\rho) \sim \rho^{1+\alpha}$ with $\alpha = 0.17$. Solid line is our model fit.  In both diffusion models, our social-tie density model fits better with mean square error 29\% and 41\% less than the power-law fit respectively.}}
\label{figdif}
\end{figure}
\end{landscape}

To test the hypothesis that a city's productivity is related to how far information travels and how fast its citizens gain access to innovations or information, it is natural to examine how this information flow scales with population density, and to quantify the functional relationship between link topology and information spreading.  We therefore simulated two models of contagion of information diffusion~\cite{kermack1150,anderson1991infectious, centola2007} on networks generated by our model.  The first contagion model simulates diffusion of simple facts, where a single exposure is enough to guarantee transmission.  The second more complex diffusion model is typical of behavior adoption, where multiple exposures to a new influence/idea is required before an individual adopts it.  In Fig. \ref{figdif} we see that that both diffusion models generate the same scaling of information diffusion rate. As a consequence we conclude that an explanation for the observed super-linear scaling in productivity with increasing population density is the super-linear scaling of information flow within the social network.

\begin{figure}[h]
\centering
\includegraphics[width=\textwidth]{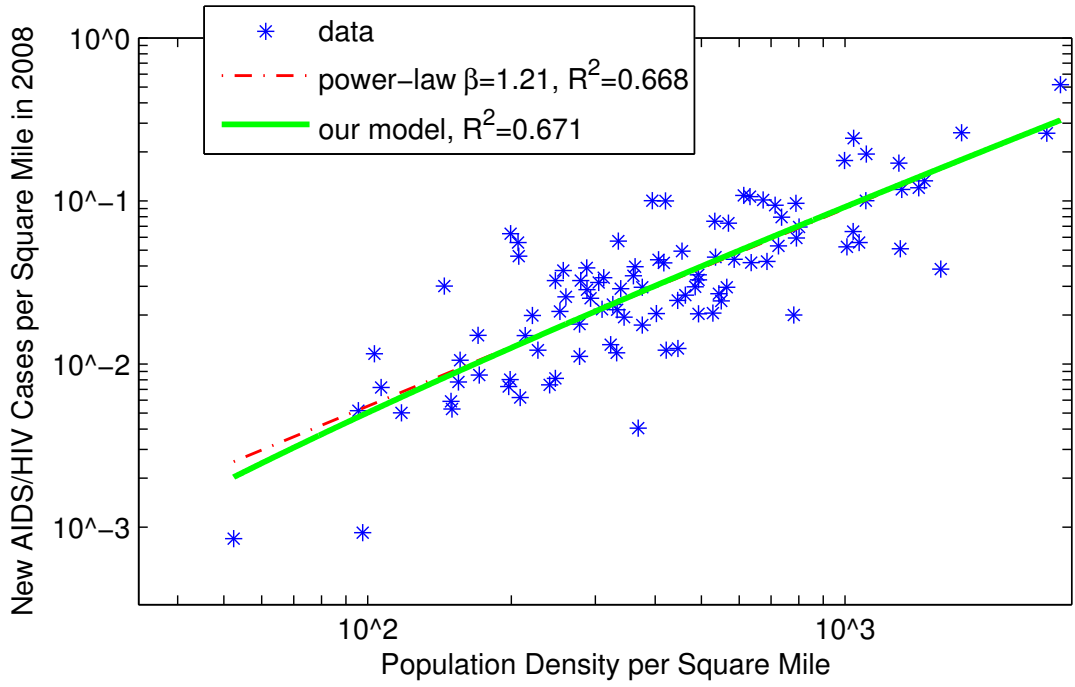}
\caption{\label{fighiv}This figure illustrates the relationship between density and HIV spreading rate in US MSAs from recent CDC and US Census datasets. As we expected, density plays an important role in describing the super-linear growth pattern, and our model fits reasonable well the real data.}
\end{figure}

\paragraph{Population-level variables} As a test case for our hypothesis, we study the prevalence of HIV infections in cities in the United States.  In Fig. ~\ref{fighiv}, we plot the prevalence of HIV in 90 metropolitan areas in 2008 (data sourced from United States Center for Disease Control and Prevention (CDC) reports and the 2010 US Census) as a function of population density. As the figure indicates, there is fairly good agreement between the data and the curve generated by our model of diffusion. 
 
\begin{landscape}
\begin{figure}
\centering
\includegraphics[width=23cm]{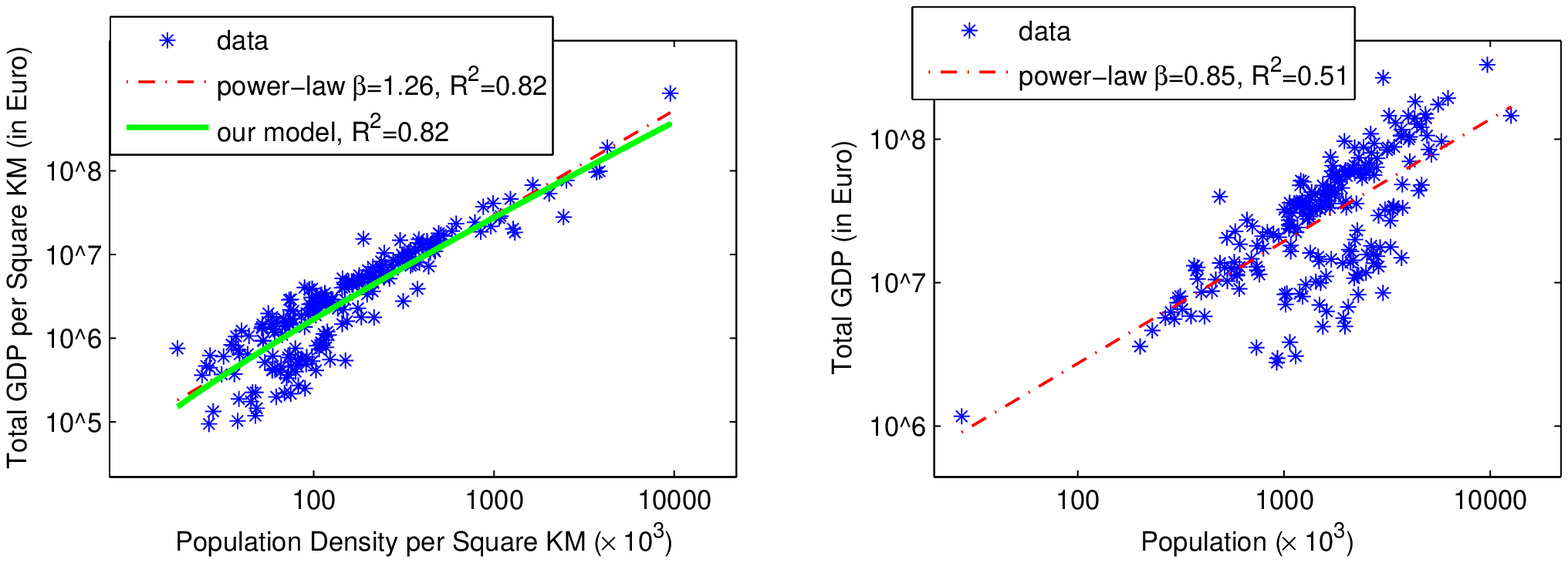}
\caption{Here we demonstrate the correlation between GDP, population and population density for all NUST-2 regions in EU. The left figure shows the correlation between density and GDP, which suggests a strong super-linear growth correlation. Both our model and a power-law fit can capture the super-linear relationship; the right figure shows the correlation between population and GDP with negative correlation.}
\label{fig:eurogdp}
\end{figure}
\end{landscape}

The same agreement holds for European cities on economic indicators. In Fig. \ref{fig:eurogdp}, we plot the overall GDP per square km in NUST-2 (Nomenclature of Territorial Units for Statistics level-2) regions in the EU as a function of population density $\rho$ as well as population size.  The NUST-2 regions are defined by the EU as the city-size level territorial partition for census and statistics purposes~\cite{european2003regulation}. We find a strong positive correlation between density and the corresponding urban metric with a super-linear scaling component, but conversely a much weak and sub-linear growth pattern on raw population size. While not the main focus of this paper, we show that the super-linear growth on density can be often be indicated in data as super-linear growth on population, and that density is a better indicator for socio-economical growth than population--see Supplementary Information, Section S2. 

Note that in both datasets the scaling exponents are restricted within a narrow band $1.1 \le \beta \le 1.3$, potentially suggesting a common mechanism behind both the prevalence of HIV and scaling of GDP with respect to the population density.  An advantage afforded by our model is the need to dispense with parameter tuning, as the model naturally produces this scaling within a reasonable margin of error.  Thus, by considering social structure and information/disease flow as a major driving force in many of the city indicators, our approach provides a unique and general theory to the super scaling phenomena of cities.

\vspace*{10px}
\noindent \textbf{Discussion}
\vspace*{10px}

\noindent In this paper we propose social tie density (the density of active social ties between city residents) as a key determinant behind the global social structure and flow of information between individuals. Based on this we have described an empirically grounded generative model of social tie density to account for the observed scaling behavior of city indicators as a function of population density. 

The model predicts that social tie density scales super-linearly with population density, while naturally accounting for the narrow band of scaling exponents empirically observed across multiple features and different geographies. We note that this is achieved without the need to recourse to parameter tuning or assumptions about modularity, social hierarchies, specialization, or similar social constructs.  We therefore suggest that population density, rather than population size per se, is at the root of the extraordinary nature of urban centers. As a single example, metropolitan Tokyo has roughly the same population as Siberia while showing remarkable variance in criminal profile, energy usage, and economic productivity. We provide empirical evidence based on studies of indicators in European and American cities (both categories representing comparable economic development), demonstrating that density is a superior metric than population size in explaining various urban indicators. 

Our argument suggests that the reasons for creating cities are not that different from creating work environments like research institutions. While current technology makes remote communication and collaboration extremely easy and convenient, the importance of packing people physically close within each other is still widely emphasized~\cite{eagle2009inferring,wu2008pentland,pentland2008honest}. We argue that cities are operating under the same principle---as a consequence of proximity and easy face-to-face access between individuals, communication and ultimately productivity is greatly enhanced. 


We of course note certain caveats and limitations of our study. The density of social ties is intrinsically a function of the ease of access between residents living in the same city. Consider the example of Beijing in China, which has a very high population density. Due to its traffic jams, Beijing currently is de-facto divided into many smaller cities with limited transportation capacities between them and consequently may not demonstrate a higher social tie density than other cities with a much lower population density. Thus a direct comparison of the model predictions with a similarly dense area such as Manhattan is not feasible.

The same limitation applies to a horizontal comparison between cities at different levels of economic development.  A  large city in Uganda may demonstrate comparable tie density with Manhattan. However the vast differences in education, infrastructure, political stability among others, naturally leads to different economic health of these two cities. Thus in our analysis, we chose to compare cities within the United States and the European Union such that these extraneous factors are controlled for.  Thus currently, the validity and efficacy of our model can only be justified within this type of controlled horizontal comparison. 

A number of theories of urban growth suggest the importance of specialist service industries, or high-value-add workers, as generative models of city development. While our model does not disprove these theories, it provides a plausible and empirically-grounded model that does not require the presence of these special social structures.  The other theories must therefore appeal to different sorts of data in order to support their claims.  Cities are one of most exceptional and enduring of human inventions. Most great cities are exceptions in their own right: a New Yorker feels out of place in Los Angeles, Paris, or Shanghai.  However, this exceptionalism may be more due to our attention to human-scale details than to the underlying structures.  In this paper we have presented a generative theory that accounts for observed scaling in urban growth as a function of social tie density and the diffusion of information across those ties.  It is our hope that this provides both a foundation for the commonalities across all cities and a beginning point for which divergence between specific cities can be explored.

\vspace*{10px}
\noindent \textbf{Acknowledgements}
\vspace*{10px}

We thank L. M. A. Bettencourt and C. A. Hidalgo for their insightful comments and help on the manuscript. 

\clearpage

\clearpage

\clearpage

\clearpage

\clearpage

\bibliographystyle{unsrt}

\bibliography{scibib}

\clearpage

\newpage

%



\noindent{\Large Supplementary Information for\\[0.5cm]Urban characteristics attributable to density-driven tie formation}

\noindent\textit{}\\
\noindent
    Wei Pan,
    Gourab Ghoshal,
    Coco Krumme,
    Manuel Cebrian,
    and Alex Pentland

\renewcommand*\contentsname{Table of Contents}
\tableofcontents
\listoffigures
\clearpage

\begin{landscape}
\begin{figure}[t]
\centering
\includegraphics[width=23cm]{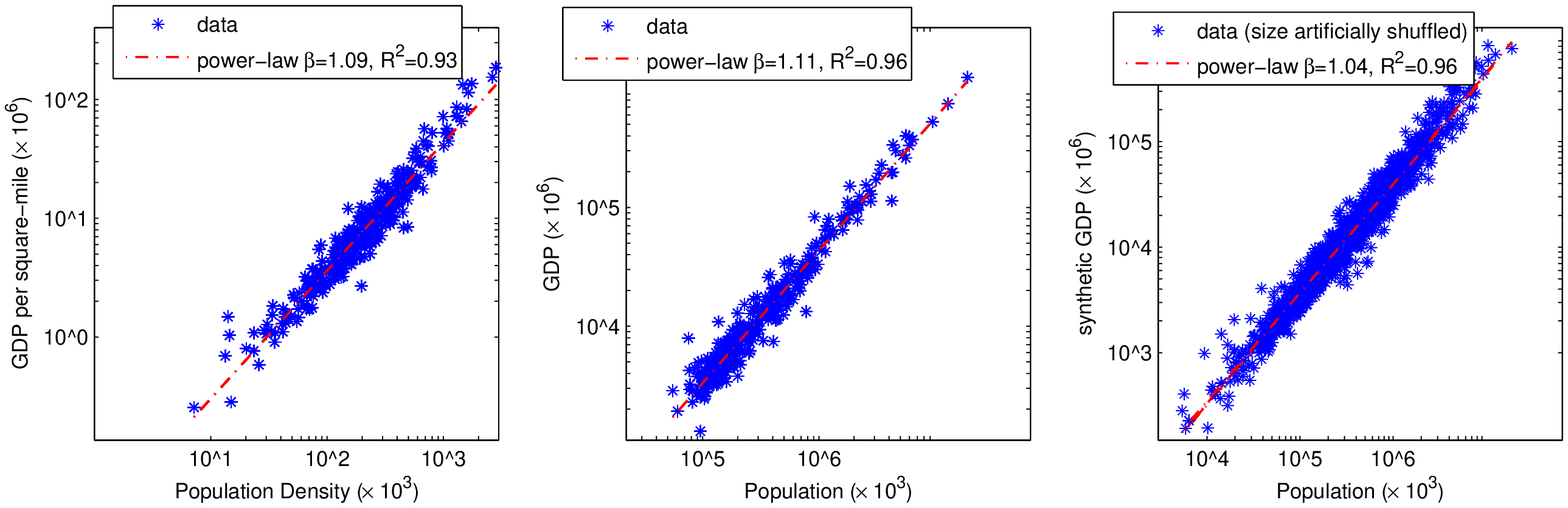}
\caption[The rescaled Gross Domestic Product (per square mile) aggregated for all MSA's in the United States for the year 2008, plotted as a function of the population density.] {Left panel: The re-scaled Gross Domestic Product (per square mile) aggregated for all MSA's in the United States for the year 2008, plotted as a function of the population density $\rho$. Middle panel: the GDP now plotted as a function of the population size. In both cases we find a clean dependence on the (re-scaled) GDP on population density as well as the population size, with a power-law fit to the form:  $x \sim y^{\beta}$---where $x$ is GDP and $y$ is population (density)---yielding an exponent $\beta \approx 1.1$ Right panel: We sample uniformly from the empirical distribution of MSA's and plot a ``synthetic" GDP---GDP per unit area multiplied by the sampled population size---as a function of population size finding once again a similar scaling relation. This appears to suggest that the empirically observed correlation between GDP and population size may in fact just be an artifact of the correlation between population density and the rescaled GDP.}
\label{figshuffled}
\end{figure}
\end{landscape}

\clearpage

\begin{landscape}
\begin{figure}[t]
\centering
\includegraphics[width=23cm]{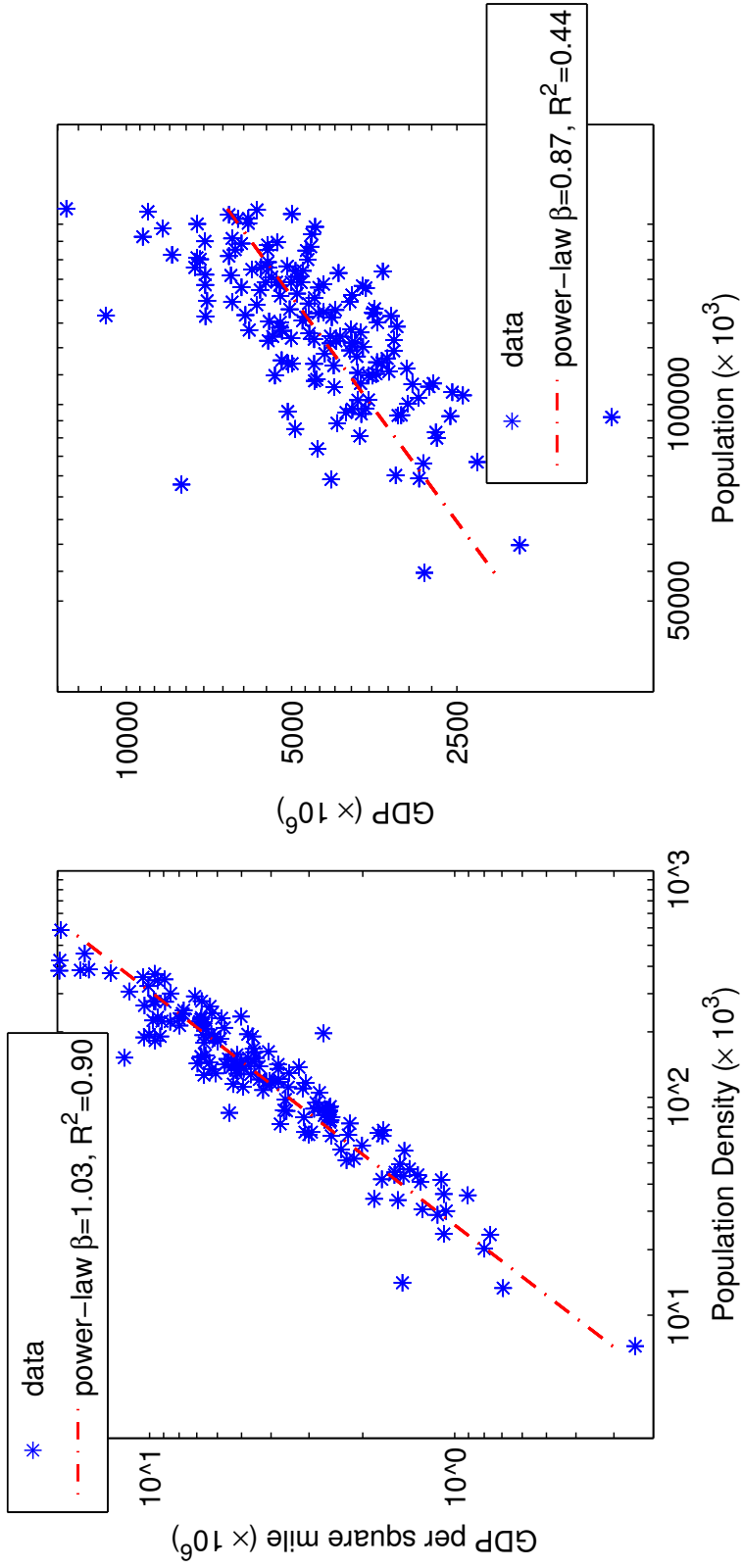}
\caption[The rescaled GDP as a function of density for cities with smaller populations.]{Left panel: The rescaled GDP as a function of density for cities with smaller populations. Right panel: the corresponding plot for GDP as a function of population. In this end of the population scale, we find that density continues to correlate strongly with GDP, whereas the correlation is far less apparent for raw population size.}
\label{figgdpsmallpop}
\end{figure}
\end{landscape}

\clearpage

\begin{landscape}
\begin{figure}[t]
\centering
\includegraphics[width=23cm]{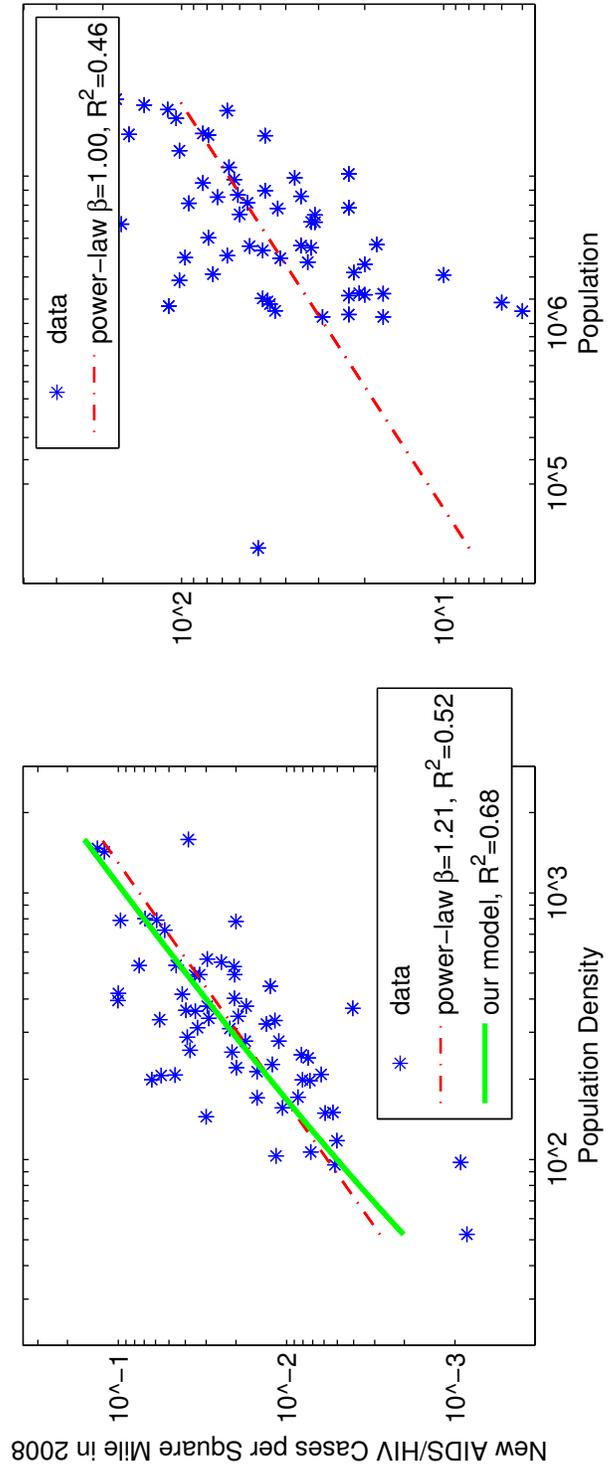}
\caption[The same as in Fig.~\ref{figgdpsmallpop} but now for new cases of HIV in the MSA's.]{The same as in Fig.~\ref{figgdpsmallpop} but now for new cases of HIV in the MSA's. While in this case the correlation with both density and population is weak, nevertheless density seems to be more important than population with respect to disease spreading.}
\label{fighivsmallpop}
\end{figure}
\end{landscape}

\clearpage
\begin{figure}[t]
\centering
\subfigure{
\includegraphics[scale=0.6]{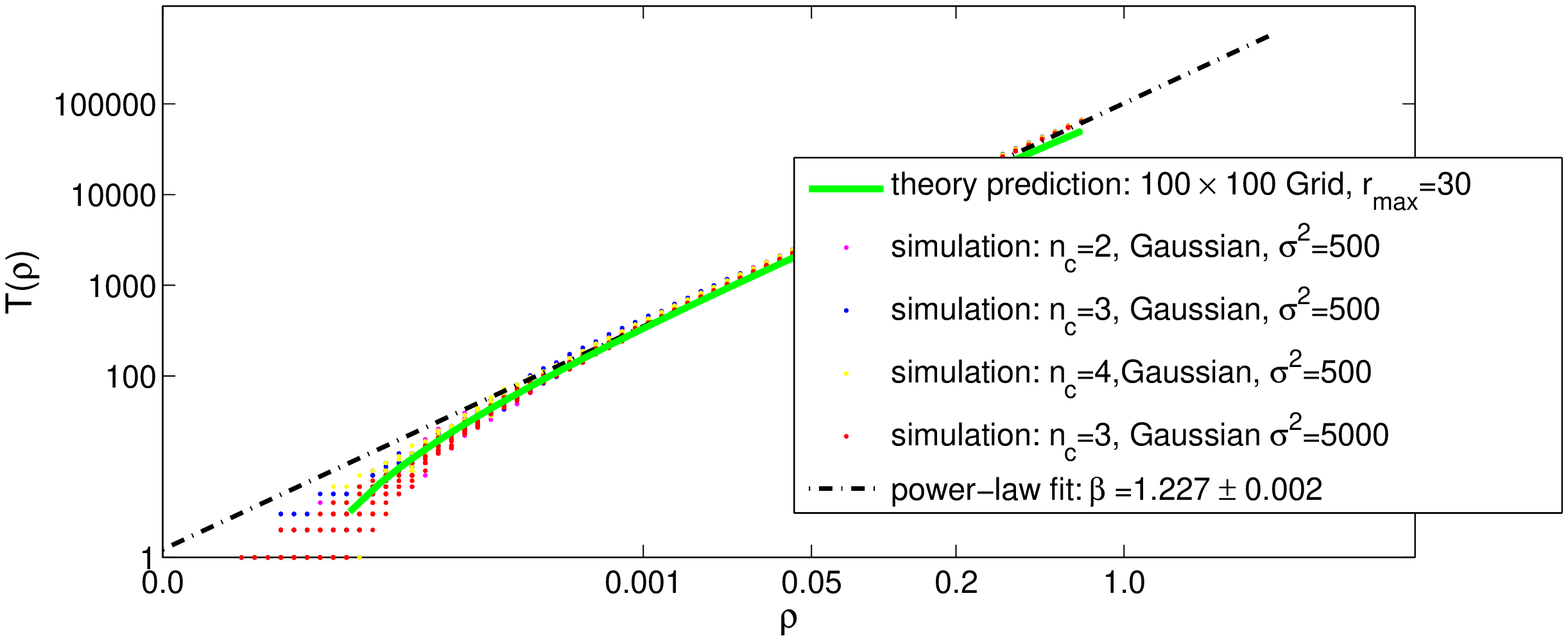}
\label{fig:uneven}
}
\subfigure{
\includegraphics[scale=0.6]{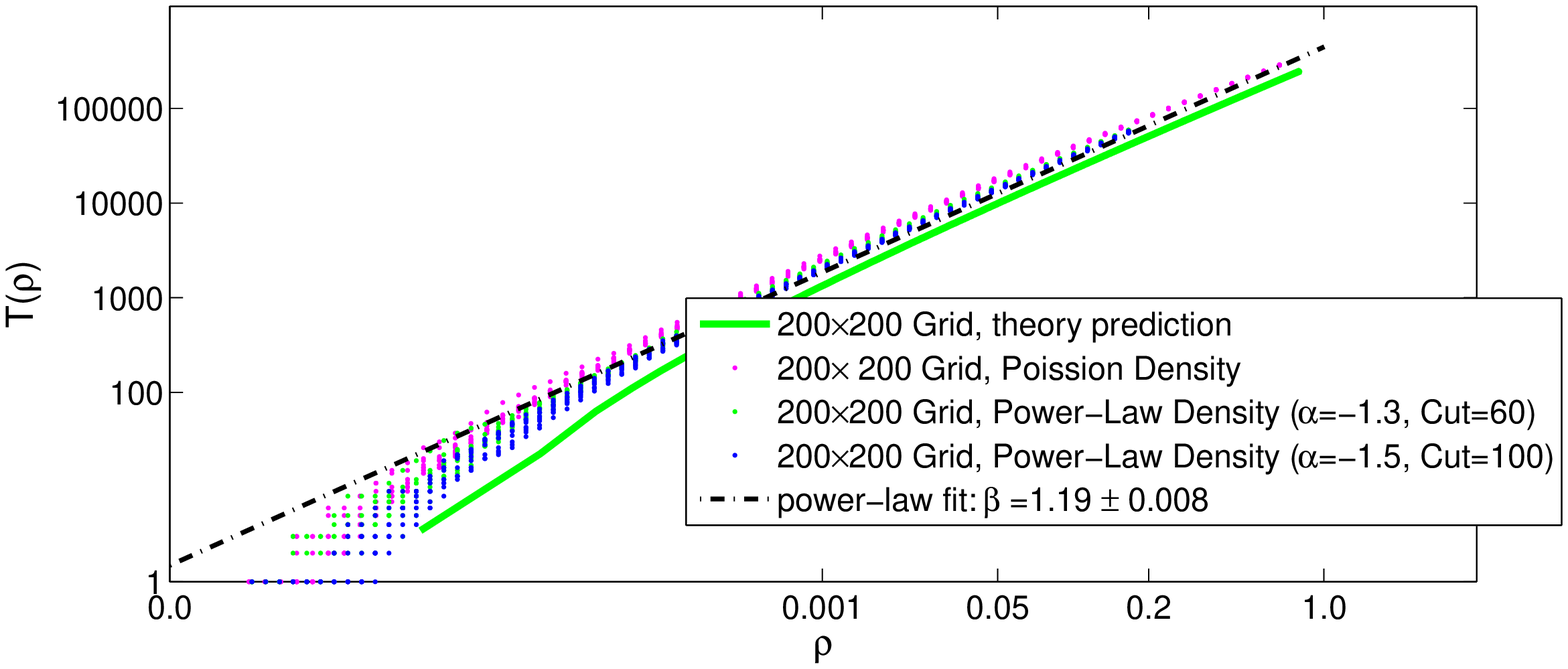}
\label{uneven_poisson}
}
\caption[The number of ties $T(\rho)$ plotted as a function of $\rho$ for various choices of non-uniform population densities.] {The number of ties $T(\rho)$ plotted as a function of $\rho$ for various choices of non-uniform population densities. The points represent the average over 30 realizations of the simulation described in the text. The top panel shows the results for placement of the nodes according to a Gaussian distribution for the densities with different numbers of city centers $c_i$ and variances $\sigma^2$. The bottom panel shows the same, but now for Poisson and power-law density distributions. It seems that the scaling for the number of ties for different choices of density distributions is well described by $T(\rho) \sim \rho \ln \rho$, while the best fit to the form $T(\rho) \sim \rho^{\beta}$ continues to yield $\beta \approx 1.2$. Thus, our analysis is robust to different (reasonable) choices for the density distribution.}
\label{nonuniform}
\end{figure}

\clearpage
\begin{figure}[t]
\centering
\includegraphics[width=\textwidth]{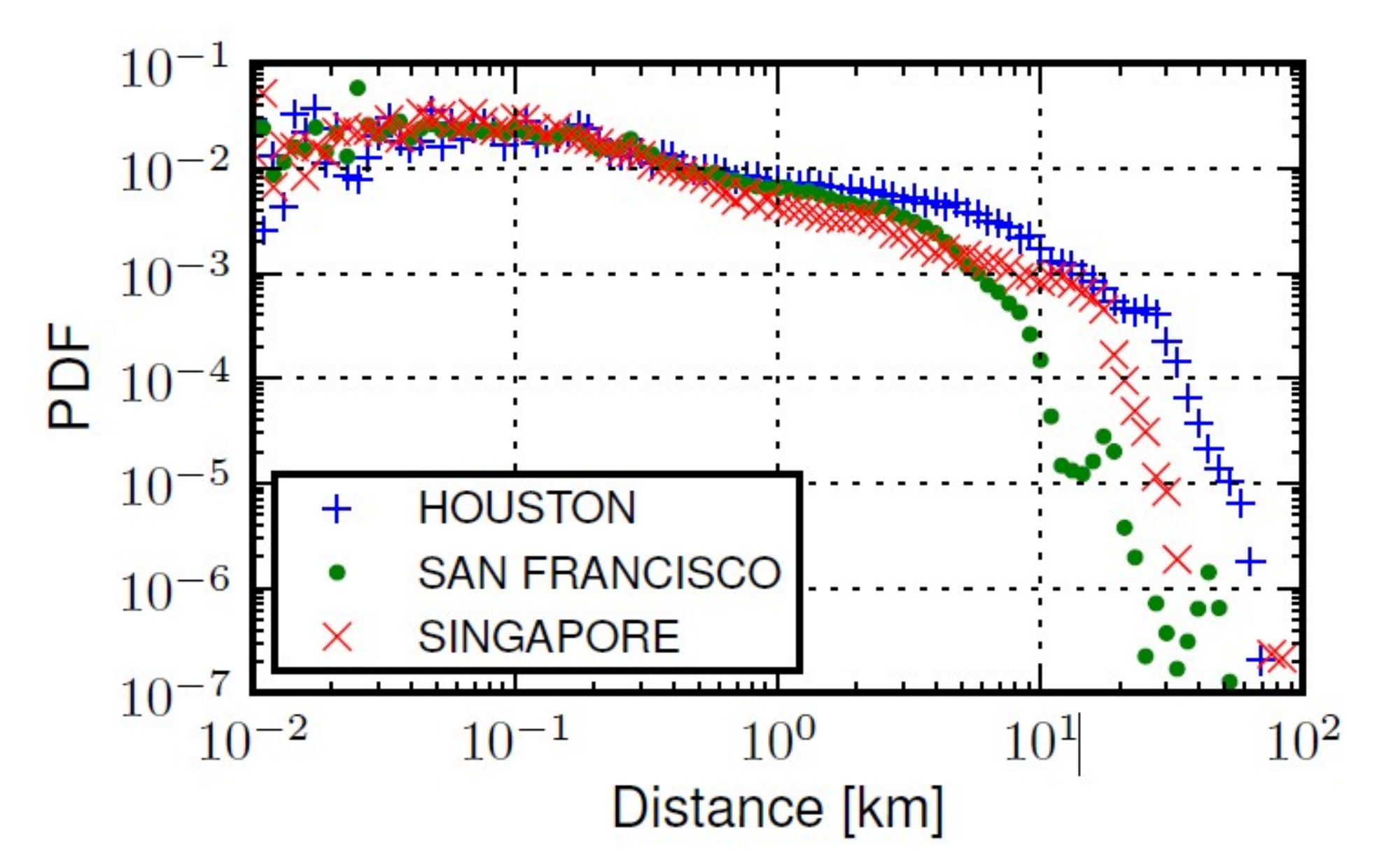}

\caption[The probability distribution function for the displacement (movement range) for city-dwellers.]{The probability distribution function for the displacement (movement range) for city-dwellers. As can be seen the distribution is flat within a cut-off threshold and decays exponentially afterwards. It is conjectured that this threshold is a natural product of urbanization~\cite{noulas2011tale}.}
\label{fig:foursquare}
\end{figure}

\clearpage
\begin{figure}[t]
\centering
\includegraphics[width=\textwidth]{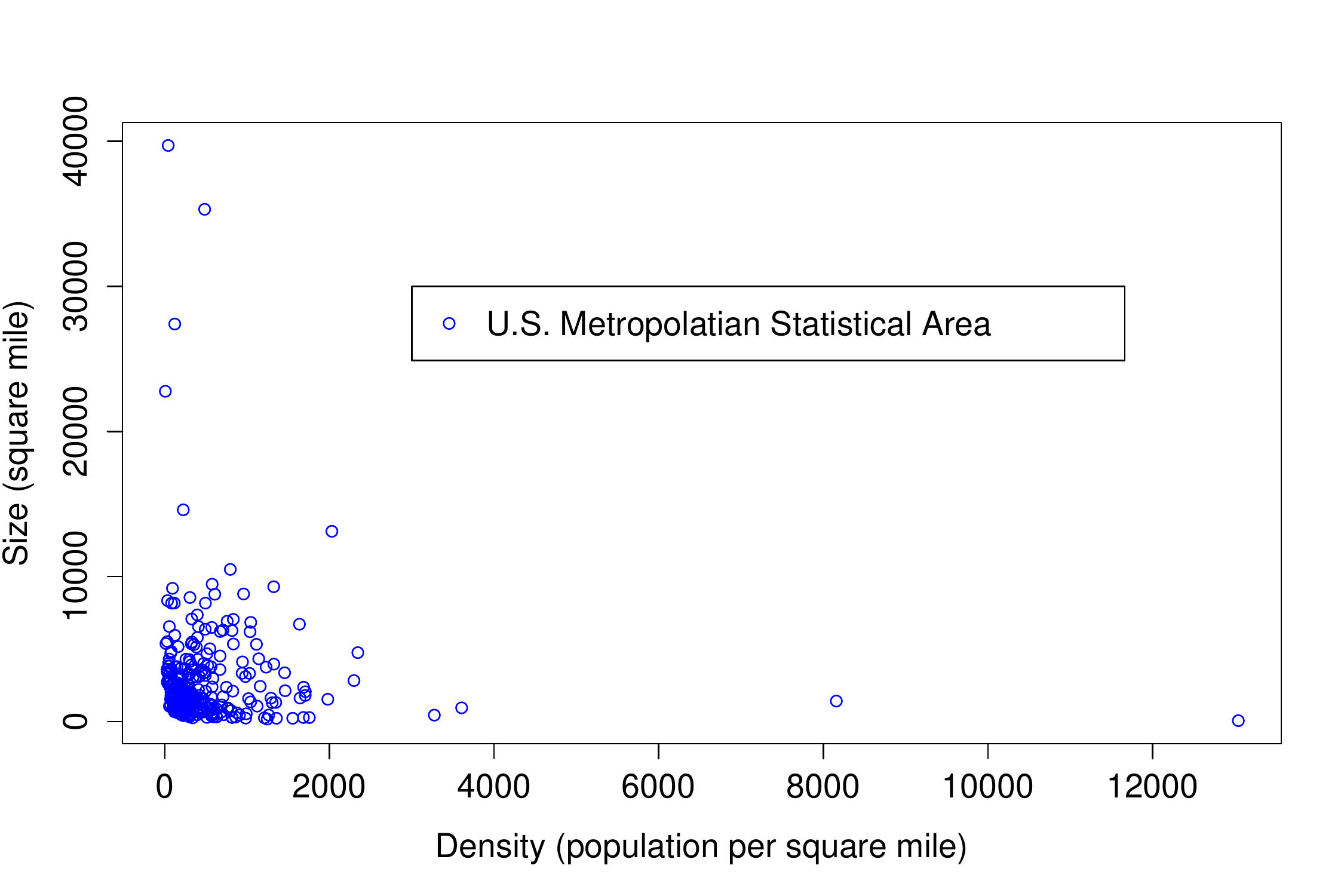}
\caption[The population density as a function of the area-size for all MSA' s  in the United States] 
{The population density as a function of the area-size for all MSA' s  in the United States. While there are a few outliers with considerable high density or large size, no trend can be seen for the majority of the data and indeed a GLM analysis yields no statistically significant correlation between density and area-size.}
\label{msadensity}
\end{figure} 

\clearpage
\begin{figure}[t]
\centering
\includegraphics[width=\textwidth]{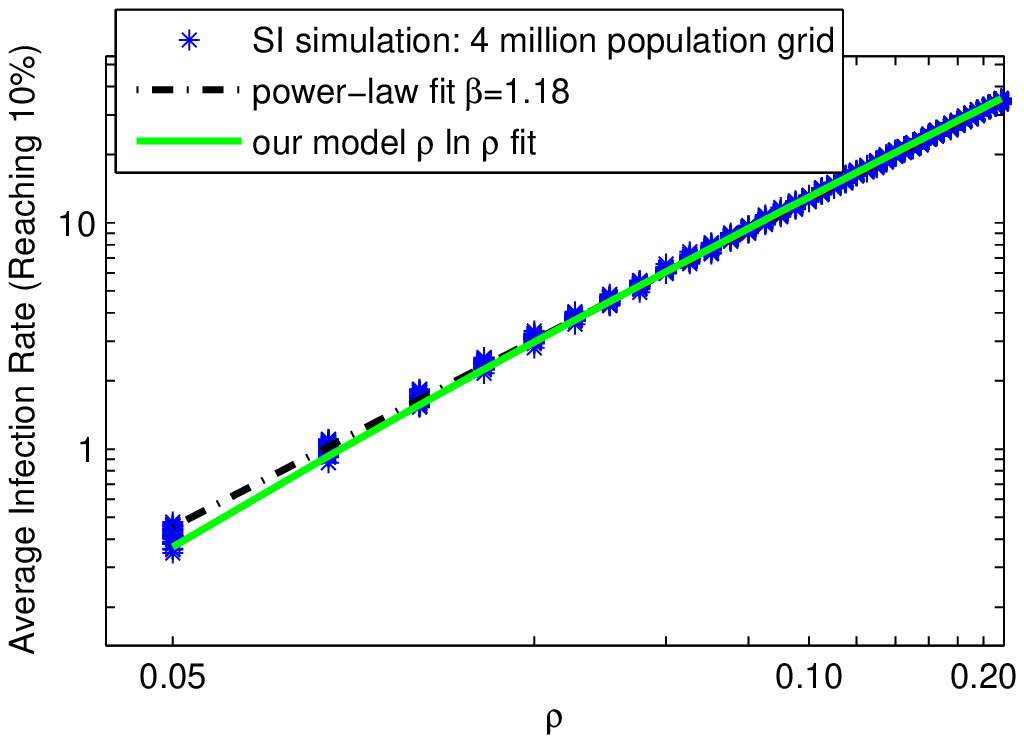}
\caption[The mean spreading rate $R(\rho)$ (Eq.~\ref{spreadspeed}) as a function of the density $\rho$  under the S-I diffusion mechanism.]{The mean spreading rate $R(\rho)$ (Eq.~\ref{spreadspeed}) as a function of $\rho$ . The points correspond to a average over 30 realizations of simulations of the SI model on a $200\times200$ grid. The dashed line corresponds to a fit of the form $R(\rho) \sim \rho^{1+\alpha}$ with $\alpha = 0.18$. The solid line is a fit to the form $\rho \ln \rho$, our social-tie density model. The simulation results match well with the empirical data for disease spreading  ($\beta=1.18$). Though not shown here, the goodness of fit for the social-tie density model far outperforms the power-law.}
\label{figsi2}
\end{figure}

\begin{figure}[t]
\centering
\includegraphics[width=\textwidth]{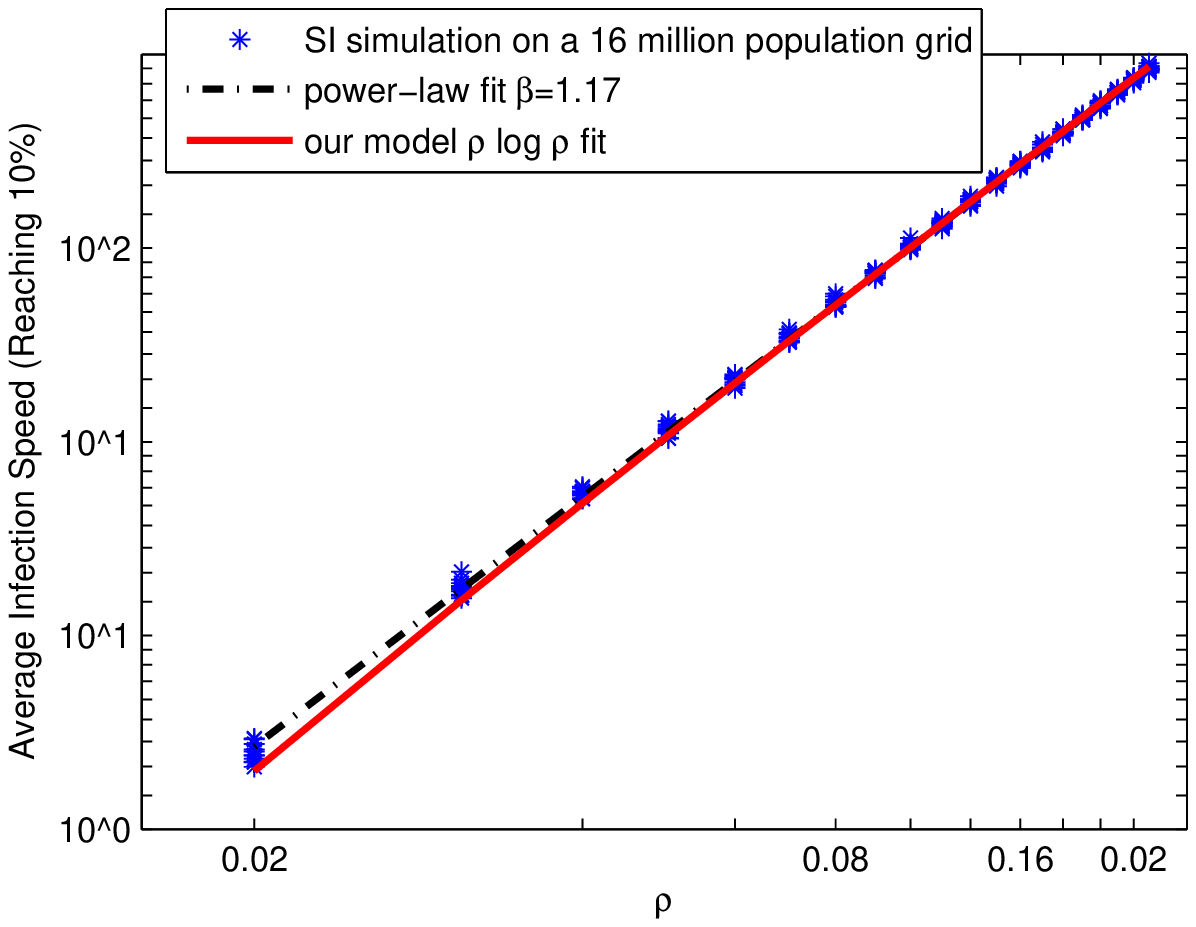}
\caption[The mean spreading rate $R(\rho)$ (Eq.~\ref{spreadspeed}) as a function of the density $\rho$ under the complex contagion diffusion mechanism.]{Similar to the above plot, this plot shows the mean spreading rate $R(\rho)$ (Eq.~\ref{spreadspeed}) as a function of $\rho$  under the complex contagion diffusion model: The points correspond to an average over 30 realizations of simulations of the complex contagion model on a $400\times400$ grid. The dashed line corresponds to the power-law fit of the form $R(\rho) \sim \rho^{1+\alpha}$ with $\alpha = 0.17$. The solid line is a fit to the form $\rho \ln \rho$, our social-tie density model, which fits better with mean square error 41\% less than the power-law fit.}
\label{figsicc}
\end{figure}

\clearpage

\begin{landscape}
\begin{figure}[t]
\centering
\includegraphics[width=23cm]{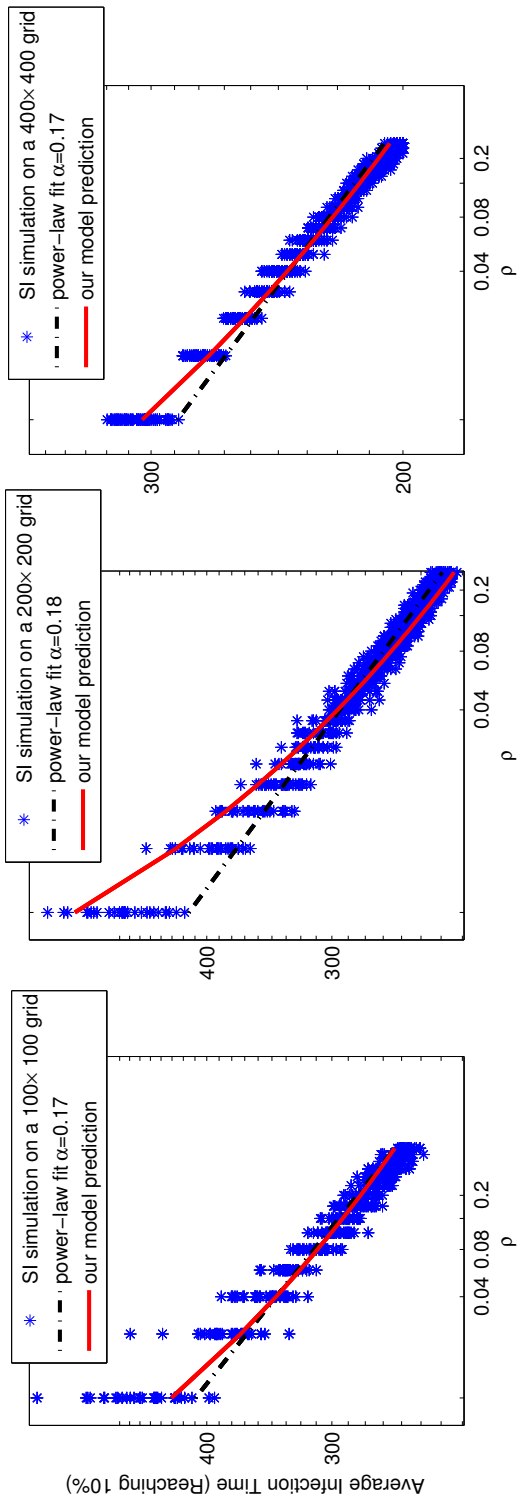}
\caption[The average infection time $S(\rho)$ (defined as the time taken to infect $10\%$ of the population) plotted as a function of $\rho$, for different values of grid size $N$.]{The average infection time $S(\rho)$ (defined as the time taken to infect $10\%$ of the population) plotted as a function of $\rho$, for different values of grid size $N$. Points correspond to the average over 30 realizations of simulations of the SI model. A fit to the form $S(\rho)\sim \rho^{-\alpha}$ (dashed line) yields $\alpha \approx 0.2$ in each of the figures. The solid line shows the corresponding fit to Eq.~\eqref{eq:logspread}.}
\label{figsi}
\end{figure} 
\end{landscape}

\clearpage
\begin{table}[h]
\caption{Growth factors $\beta$ for some urban economic factors~\cite{bettencourt2007growth}.}
\label{tab}
\begin{center}
  \begin{tabular}{ c | c  }
    \hline
    Urban Economic Indicator & Growth Factor $\beta$ \\ \hline
    New Patents & 1.27 \\                                                                                                                                    
    GDP & 1.13  - 1.26   \\                                                                                                                    
    R\&D Establishment & 1.19 \\                                                                                                                             
    Intra-city call time & 1.14 \\                                                                                                                             
    New AIDS Cases & 1.23  \\     \hline

  \end{tabular}
\end{center}
\end{table}

\clearpage

\section{Superlinear scaling of urban indicators}
Recent empirical evidence points towards a consistent scaling relation between various urban indicators and population size resolved temporally.  Bettencourt \etal~\cite{bettencourt2007growth,bettencourt2010urban,bettencourt2010unified,calabrese2011connected} have studied the relation between many urban economic indicators in a city and the population, and report a common scaling behavior of the form
\begin{equation}
Y(t) \sim N(t)^{\beta},
\label{eq:sl1}
\end{equation}
where $Y(t)$ is some urban economic indicator, and $N(t)$ is the population size at time $t$. They find that many urban indicators, from disease to productivity, grow with surprisingly similar values for the exponent $1.1 \le \beta \le 1.3$ as shown in Table \ref{tab}.  They suggest that such a scaling pattern reflects quantities such as information, innovation and wealth creation and conjecture that these are intrinsically related to social capital, crucial to the growth and sustainability of cities. While such findings, viz. the qualitative dependence of economic indicators on the population size, potentially have a profound impact---implying that global urbanization is very efficient and a key driver of economic development---there is some debate as to which is underlying mechanism as well as the precise functional relationship between the two. For instance, Shalizi~\cite{shalizi} re-examined the same dataset and suggested that the scaling relation between the two may better be explained by a logarithmic dependence rather than a power-law between the indicators---an observation consistent with the results presented in the manuscript.


\section{Density and population}

Empirical evidence suggests a consistent scaling relationship between both the population and population density as a function of urban indicators. We argue here that the scaling relation between an urban indicator and population size may in fact be an artifact of the correlation between population density and the said metric. 
 
 In Fig. \ref{figshuffled}, (middle panel), we plot the Gross Domestic Product---aggregated over all Metropolitan Statistical Areas (MSA's) in the United States measured in 2008---as a function of the population size. In the left panel we now plot the rescaled GDP (defined as the GDP per  unit area) as a function of population density. In both cases we find a super-linear scaling with an exponent $\beta \approx 1.1$. Next, we sample uniformly from the empirical distribution of MSA's and in the right panel, plot the ``synthetic" GDP---GDP per unit area multiplied by the sampled population size---as a function of population size finding once again a similar scaling relation. This appears to suggest that the empirically observed correlation between GDP and population size may in fact just be an artifact of the correlation between population density and the rescaled GDP.

There may be two  contributing factors to this phenomenon, namely the relative homogeneity in the size of MSA's 
($\sigma_{size}=2900$ sq. mi. vs $\sigma_{pop.}=1.6^6$) , coupled with the fact that  the actual variance in city sizes is ``averaged" over when plotting on a logarithmic scale. To account for this, in Fig.~\ref{figgdpsmallpop}  we re-plot the left and middle panels for Fig.~\ref{figshuffled} only for cities with smaller populations (this sample has a higher variance in size). We find that density continues to plays a strong role in GDP growth, while the correlation with population is far less apparent. 
In Fig. \ref{fighivsmallpop}  we show the corresponding plot for new HIV cases in these MSA's and find once again that density has a higher correlation with disease spread than population.

\section{The choice of city boundary $r_{max}$ 
\label{sec:rmax}}
One of the simplifying assumptions we made in our model, is that the city boundary $r_{max}$ is independent of the population density. Here we provide supporting empirical evidence for our proposition. 

 
Empirical measurements of FourSquare location data  has shown that physical mobility boundaries~\cite{noulas2011tale} exist in cities (see Fig.~\ref{fig:foursquare}). In other words, the range distribution of physical movements and activities for humans living in a city is flat within a threshold distance, and decays exponentially above the threshold.

 
Next, we compare the size of a Metropolitan statistical area (MSA)---defined by counting adjacent areas tied to urban centers socioeconomically, with the area size itself implying an underlying movement pattern $r_{max}$~\cite{usmsa}---to the population density in those MSA's.
In Fig. \ref{msadensity} we plot the size of the different MSA's as a function of the population density (the data is taken from the US census in the year 2000). A Generalized Linear Model regression yields no correlation ($p > 0.50$) between the population density and city size. 



\section{Model Description}
We propose to model the formation of ties between individuals (represented as nodes) at the resolution of urban centers. Since our model is based on geography a natural setting for it is a 2D Euclidean space with nodes denoted by the coordinates $\vec{x}_i \in {\mathbf R}^2$ on the infinite plane.  Furthermore, we also assume that these nodes are distributed uniformly in the space, according to a density $\rho$ defined as,
\[\rho = \text{\# nodes per unit area}. \] 
Following~\cite{liben2005geographic}, the probability to form ties between two nodes in the plane will be according to 
\begin{equation}
P_{ij} \propto \frac{1}{\mathrm{rank}_i (j)},
\label{eq:probij}
\end{equation}
where the rank is defined as:
\begin{equation}
\mathrm{rank_i (j)}:= \left|\{k:d(i,k) < d(i,j)\}\right|,
\label{eq:rank}
\end{equation}
with $d_{ij}$ the Euclidean distance between two nodes.
If $j$ lies at a radial distance $r$ from node $i$, then the number of neighbors closer to $i$ than $j$ is just the product of the density and the area of the circle of radius $r$, and thus the rank is simply,
\begin{equation}
\mathrm{rank_i}(j) = \rho \pi r^{2},
\label{eq:rankcont}
\end{equation}
and thus the probability for an individual to form a tie at distance $r$ goes as $P(r) \sim 1/r^2$, similar in spirit to a Gravity model as empirically measured by~\cite{krings2009gravity}.
Since $P(r)$ is a probability, it is necessarily bounded in the interval $(0,1)$ and therefore there is a minimum radius $r_{min}$ defined by the condition,
\begin{equation}
\frac{1}{\rho \pi r_{min} ^2} = 1,
\label{eq:rmincond}
\end{equation}
The process now evolves as follows. At each step a new node is introduced into the plane; as it is introduced it forms ties with other nodes (if present) with probability $P(r)$, while existing ties in the city remain unchanged.
Consider a randomly chosen node in the plane, say $i$ and let us draw a circle of radius $r$ centered at $i$. The number of ties that node $i$ forms at some distance $r$ is the product of the expected number of nodes at that distance $2 \pi r dr$ and the probability of forming ties $P(r)$. Integrating over $r$, the total number of ties that $i$ is expected to form is given by,
\begin{equation}
t(\rho) = \int^{r_{max}}_{r_{min}} 2 \pi r dr \rho \times P(r) + 1,
\label{eq:noft}
\end{equation}
the additional term of $1$ accounts for the fact that $i$ necessarily forms a tie with a node within radius $r_{min}$ centered at $i$. The parameter $r_{max}$ denotes an upper cutoff for the integral to bound it and reflects the fact that $r$ has natural limits at long distances, such as at the border of a metropolitan area where geographical distance is no longer a de-equalizing force~\cite{noulas2011tale}.  Additionally, we assume that $r_{max}$ is independent of density, an assumption supported by empirical evidence (see~\cite{noulas2011tale} and Section \ref{sec:rmax}). Substituting in the appropriate terms we have,
\begin{eqnarray}
t(\rho) &=& \int^{r_{max}}_{\frac{1}{\sqrt{\pi \rho}}} 2 \pi r dr \rho \times \frac{1}{\pi r^2 \rho} + 1 \nonumber \\
&=& \ln{\rho} + C \label{i2}, 
\end{eqnarray}
where $C = 2\ln{r_{max}} + \ln{\pi} + 1$. 
To get the total number of ties formed by all nodes within an unit area we integrate over the density to get,
\begin{eqnarray}
T(\rho) &=& \int_0^{\rho} t(\varrho) d\varrho \nonumber \\
&=& \rho\ln{\rho} + C' \rho,
\label{eq:tprime}
\end{eqnarray} 
where $C'=C-1$.  Thus in the setting of our model, the number of ties formed between individuals, to leading order goes as $T(\rho) \sim \rho \ln \rho$, a super-linear scaling consistent with the observations made in~\cite{bettencourt2010unified}.




\section{Non-uniform population density distribution}
It may be argued that assuming a uniform distribution for the population may be an oversimplification of the actual densities found in cities. While an analytical treatment of the same is rather involved, to verify the robustness of our findings conditioned on our assumption (and indeed to measure the deviations of our prediction moving away from our assumption), we modify our simulations accordingly.  

As most cities seem to have a dense core (city center), or 
a series of densely populated regions (downtowns, main streets) 
interspersed by sparser areas, we consider a distribution of 
density on the grid captured by
a mixture of 2D Gaussian distributions with randomly 
selected centers with the $i$'th center denoted as $c_i$ at 
location $r_i$. Correspondingly nodes are introduced to 
the grid according a probability sampled from the sum of Gaussian distributions 

\begin{equation}
\sum_{i} f (r_i, \sigma^{2}).
\end{equation}

We run repeated simulations by both varying the number of city centers $c_i$ as well as the variance $\sigma^2$. The results are shown in Fig. \ref{nonuniform}. We find that the results are not too different from the case for uniform population density and continue to be well described by our theoretical expression~\eqref{eq:tprime}. Different choices for the population density such as Poisson or power-law, also shown in~Fig. \ref{nonuniform} produce similar results.

\section{Diffusion of Disease and Information}
To understand how the growth processes of cities work to create observed super-linear scaling, it is not sufficient to state the expected level of link formation. After all, links themselves do not create value; rather, the pattern by which links synthesize information is at the heart of value-creation and productivity. This line of investigation is beyond just academic interest, as it is well known that the structure of the network has a dramatic effect in the access to information and ideas \cite{granovetter1973strength,eagle2010network}, as well as epidemic spreading \cite{colizza2006role,wang2009understanding}. If we assume that a city's productivity is related to how fast its citizens have access to innovations or opportunities, it is natural to examine how this speed scales with population density under our model. The same analogy motivates the investigation into disease spreading: with more connectivity, pathogens spread faster, and thus it is of interest to quantify their functional relationship.  

We discover, by running a simple SI spreading model on the density-driven networks from previous simulation, and
discover that the mean diffusion speed grows in a super-linear fashion with $\beta \approx 1.2$, in line with our previous results and match well with the disease spreading indicators in cities~\cite{bettencourt2007growth}. Correspondingly, we propose that an explanation for both the super-linear scaling in productivity and disease is the super-linear speed at which both information and pathogens travel in the network with a characteristic scaling exponent.

Assuming that the spread of information and disease are archetypes of simple contagions, we run the SI (Susceptible-Infectious) model~\cite{kermack1150,anderson1991infectious} on networks generated by our model and measure the speed at which the infection reaches a finite fraction of the population.  We start by generating networks according to the process described in the previous section and then randomly pick $1\%$ of the nodes as seeds (i.e initial infected nodes). At each time step, the probability of an infection spreading from an infected to a susceptible node is denoted as $\epsilon$, which we fix at $\epsilon=1 \times 10^{-2}$. The simulation terminates at the point when $10\%$ of the populations is in the infected state.  The networks generated are snapshots at different densities $\rho$ and as before we vary both the size of the grid $N$. 

Denoting $S(\rho)$ as the number of time steps taken to infect $10\%$ of the population, the mean spreading rate $R(\rho)$ can be written as,
\begin{equation}
R(\rho) = \frac{\rho}{S(\rho)}.
\label{spreadspeed}
\end{equation}
The results of our simulations are shown in Fig.~\ref{figsi}, where we show $S(\rho)$ as a function of $\rho$. Fitting it to a form:
\begin{equation}
S(\rho) \sim c\rho^{-\alpha},
\label{spread}
\end{equation}
yields a value of $\alpha \approx 0.20$. 
Assuming that the mean spreading rate is proportional to the network density (i.e. $R(\rho) \propto T(\rho)$), we also fit the data to the form

\begin{equation}
S(\rho) \sim \frac{\rho}{\frac{1}{k}T(\rho)} = \frac{k}{\ln \rho + C'},
\label{eq:logspread}
\end{equation}
where $T(\rho)$ is the expression in Eq. \ref{eq:tprime} and $k$ is a constant. As can be seen the curve corresponds well to the data points.

In Fig. \ref{figsi2} we explicitly plot $R(\rho)$ as a function of $\rho$ and find that the curves are well fitted by the power-law with an exponent $\beta \approx 1.2$, in line with our previous results and match well with  the disease spreading indicators in cities~\cite{bettencourt2007growth}. By assuming the spreading speed is proportional to our social tie density, we plot in Fig. \ref{figsi2} our model prediction on diffusion rate, which yields an excellent fit with mean square error 29.4\% lower than the power-law fit.  Correspondingly we propose that an explanation for both the super-linear scaling in productivity and disease is the super-linear speed at which both information and pathogens travel in the network with a characteristic scaling exponent, proportional to the social tie density.

\subsection{Complex Contagion Diffusion}
In addition to the S-I model simulation, we also consider the complex contagion model~\cite{centola2007}. We assume that 10\% of the population is simple contagion: an individual will be infected by only one infected neighbor; the rest 90\% of the population is complex contagion: it takes at least two neighbors to change the individual. The rest of the simulation is identical to the simulation with the S-I model, and we count the time steps needed to infect 10\% of the population. We show the result in Fig. \ref{figsicc}.

As shown in Fig. \ref{figsicc}, we observe that $R(\rho)$ also grows super-linearly with an exponent $\beta\approx1.17$. Therefore, we confirm that under our social tie density model, both diffusion mechanisms lead to the same scaling results. We also show our logarithmic fit is better than the power-law fit, with mean square error 41\% lower, suggesting our model explains better the super-linear information travel speed in the network.

\bibliographystyle{nature}
\bibliography{scibib}

\end{document}